\documentstyle[11pt,aaspp4]{article}
\lefthead{George et al.}
\righthead{Is Shedir Variable?}
\begin{document}
\title{Is Shedir Variable?}
\author{Samuel J. George,\altaffilmark{1} Ian R. Stevens, \altaffilmark{1} Steven A. Spreckley\altaffilmark{1}}
\altaffiltext{1}{School of Physics and Astronomy, University of Birmingham, Birmingham, United Kingdom, B15 2TT}

\begin{abstract}
Before the age of modern photographic and CCD observations $\alpha$ Cassiopeiae was labelled as a variable star, though this variability has not been seen with modern instrumentation. We present an analysis of 3 years of high precision space-based photometric measurements of the suspected variable star $\alpha$ Cassiopeiae, obtained by the broad band Solar Mass Ejection Imager (SMEI) instrument on board the Coriolis satellite. Over the 3 years of observations the star appears to not show any significant variability. Also, data from the Hipparcos epoch photometry annex shows no significant variability.
\end{abstract}

\keywords{Stars: individual: $\alpha$ Cas stars: variables: general}

\section{Introduction}
$\alpha$ Cassiopeiae (RA: 00$^{\textrm{h}}$40$^{\textrm{m}}$30.5$^{\textrm{s}}$, Dec:+56$^{\circ}$32'14.5'') is the brightest star in the constellation of Cassiopeia with an apparent 
magnitude of 2.25. It has the traditional name Shedir (which may also be 
spelt as ``Shedar'', ``Shadar'', ``Schedir'', or ``Schedar''). It is a 
class K0~IIIa star (\cite{Morgan73}) at a distance of 68~pc 
(\cite{hipparcos}). In the past it has been classified as a variable star 
(\cite{Campbell}). \cite{Birt} observed that the star changes in magnitude 
from 2.2 to 2.8 but showing no clear structure to this brightening (about 
80\% of observations the star were found at the brighter limit). No 
significant variability has been detected since the 19th century. It is 
worth noting that there are  three known companions to the star have been 
listed in the Washington Double Star Catalog (\cite{washington}), though 
they are just line-of-sight optical components (the nearest companion is 
some 20'' away, and is 14th magnitude). The question - is Shedir variable was posed 
by Sir Patrick Moore in the January issue of the BBC Sky at Night Magazine whilst summarising the year ahead \cite[]{pm} and is the initial motivation behind this analysis.

In this letter we present SMEI observations of $\alpha$ Cassiopeiae with
the aim of determining if the star is variable via high precision space
based photometry. The data reduction and analysis is described in Section
2. In Section 3, we discuss the results from SMEI along with the the Hipparcos epoch photometry annex data with conclusions presented in Section 4.

\section{Data Reduction and Light-curve extraction}

The photometry we present here were obtained using SMEI. The reader is
refereed to \cite{h5} for an overview of the processing of SMEI data in
relation to the Cepheid variable star Polaris, with a more detailed
description of the data extraction and processing pipeline given in
\cite{h6}. In summary, the SMEI instrument consists of three cameras
facing respectively along, perpendicular to, and away from the Earth-Sun
direction. Each of the cameras has a field of view of 60$^\circ$ $\times$
3$^\circ$. SMEI has a roughly triangular pass band with a peak quantum
efficiency of 47\% at 700 nm and falling to 5\% at 430 nm and 1025 nm.  
Due to the satellite's orbital path the entire sky is imaged with
approximately a 100-minute cadence.  The individual 60$^\circ$ $\times$
3$^\circ$ raw images are bias subtracted, have a temperature scaled dark
current signal removed, and are flat fielded. Hot pixels and high energy
particle hits are corrected on the images via interpolation, before
aperture photometry is performed.  Aperture photometry based on a modified
from of the DAOPHOT routines (\cite{h7}) are then performed. The produced
light curves are then corrected for systematic instrumental effects.  The
data under consideration were gathered between Julian date 2452739.526 and
2453816.389 (April 2003 to March 2006). While the SMEI data extend beyond
this date, subsequent data are still undergoing processing. The full light
curve can be found in Figure~\ref{fig1} with no distinct variability seen
(a maximum point to point scatter of 0.01 mag).

\section{Results}

\begin{figure}[ht]
\figurenum{1}
\epsscale{1.0}
\plottwo{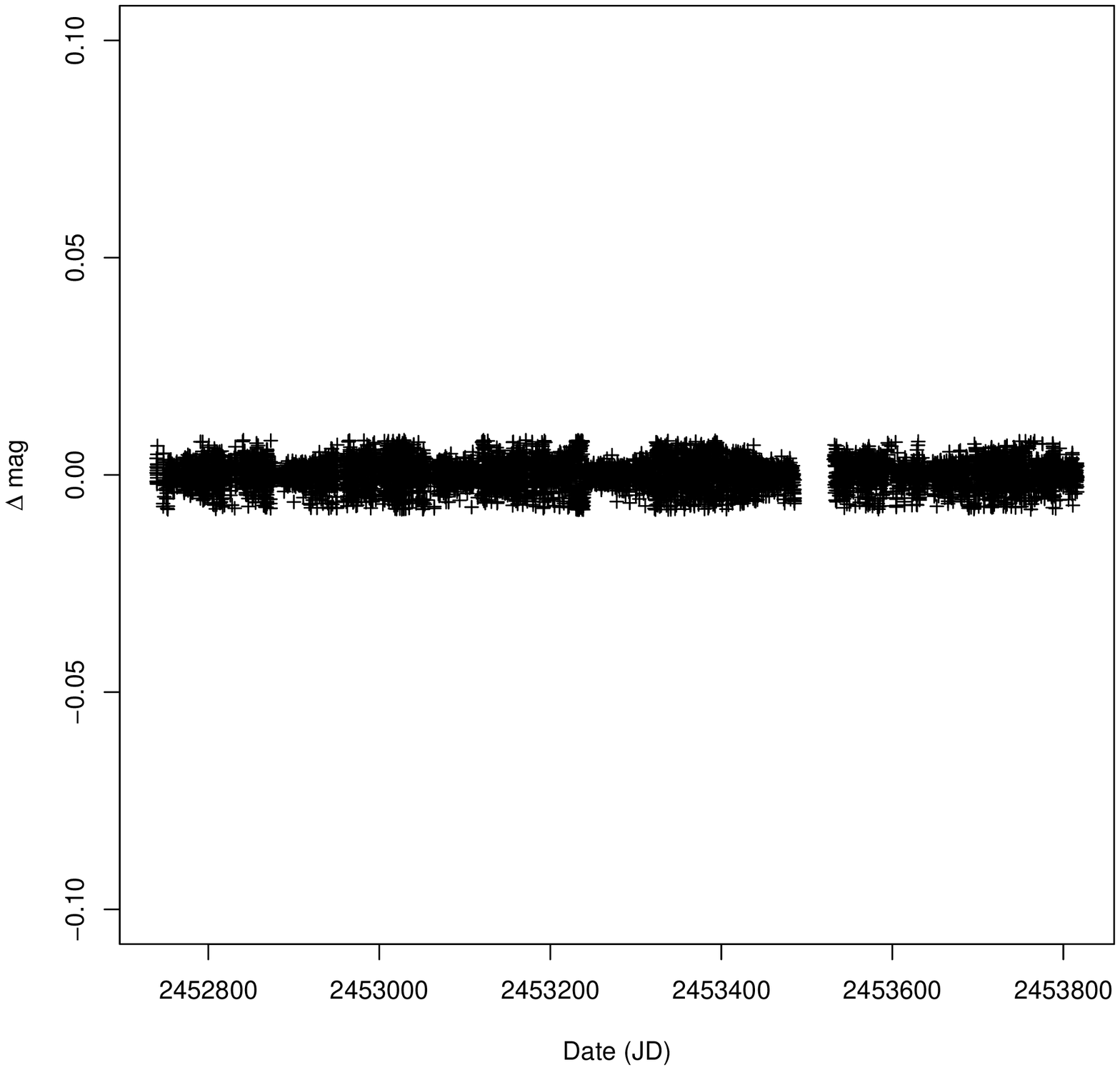}{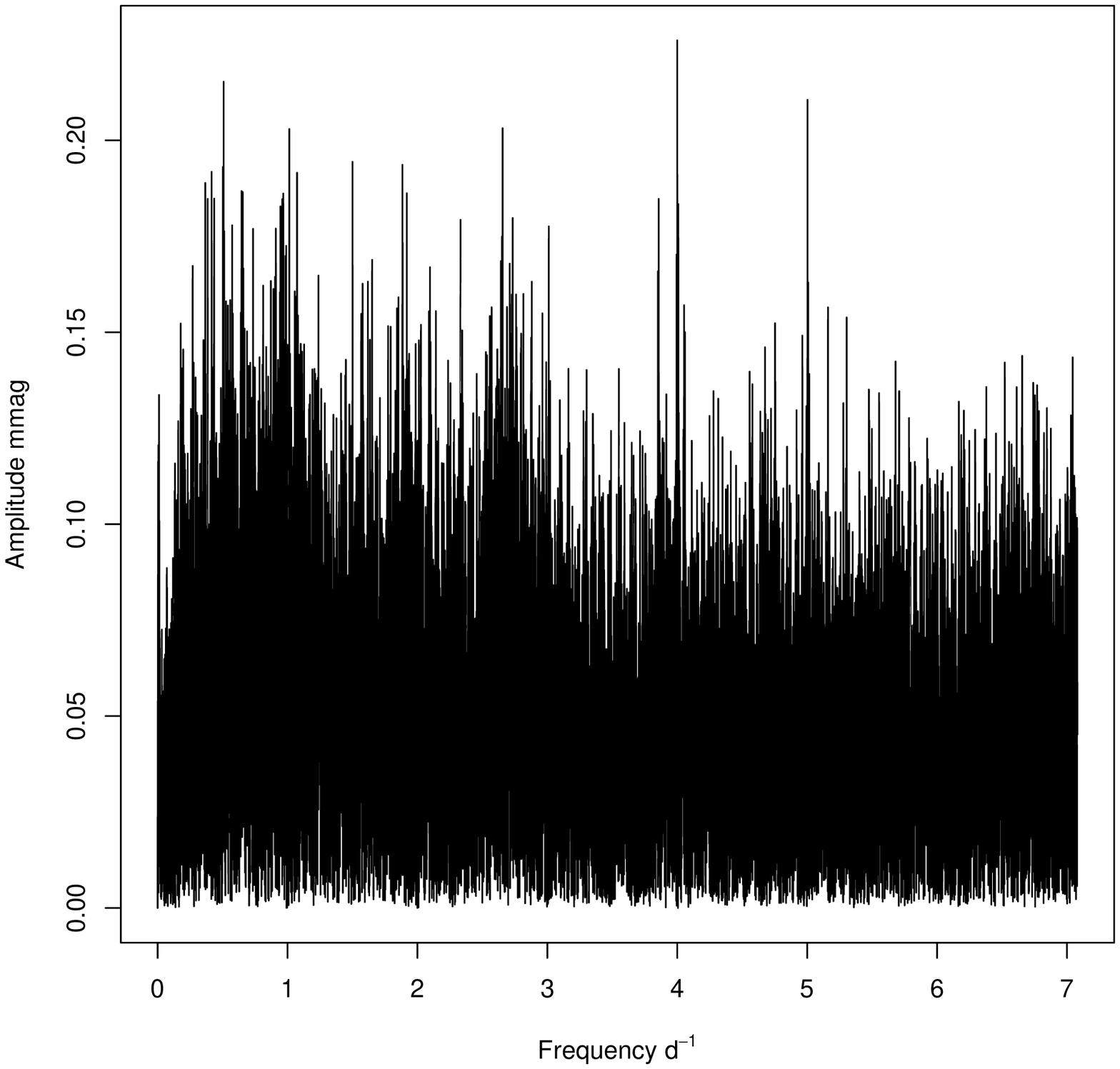}
\epsscale{0.5}
\begin{center}
\plotone{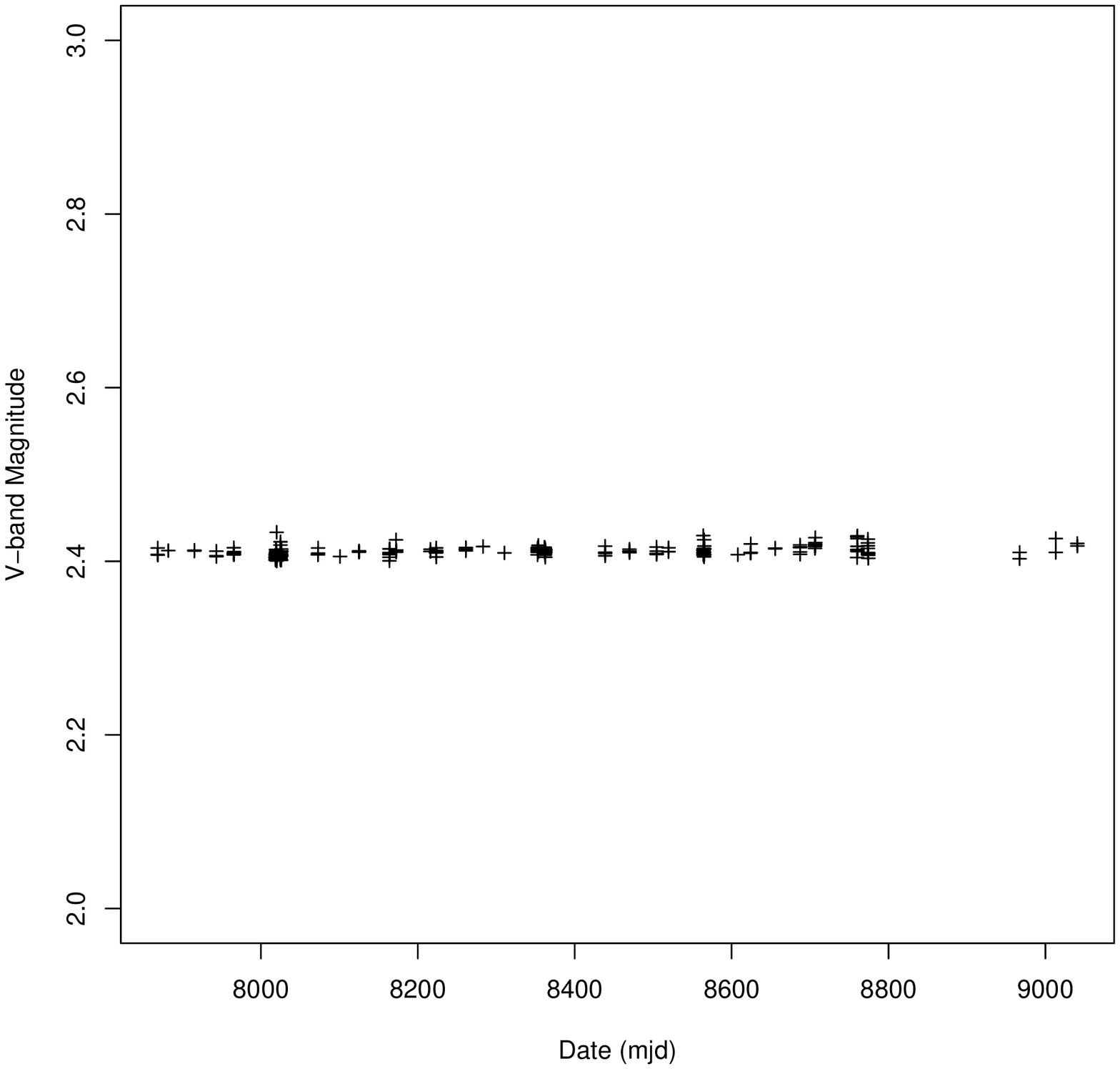}
\end{center}
\hspace*{10em}
\caption[Full light curve and FT data]{\textbf{Top Left}: The complete 3 years of the SMEI time-series showing no appreciable variability. \textbf{Top Right}: The Fourier spectrum of the SMEI data.  \textbf{Bottom}: 4 years of data obtained by the Hipparcos satellite, showing no significant variability of the source. \label{fig1}}
\end{figure}

The data was analysed for periodicites with the PERIOD04 package
(\cite{h215}). The Fourier analysis reveals features at 1 day and
harmonics of 1 day which are a feature of the SMEI data and are associated
with stray light and daily periodicites in the cosmic ray level.
Significant power is seen at 1.0, 2.0, 3.0, 4.0 and 5.0 d$^{-1}$. These
are removed using the sine-wave fitting in PERIOD04. After removal of
these features the Fourier spectrum clearly shows significant power in the
data-set and can be seen in Figure~\ref{fig1}. 

We have also used data obtained from the Hipparcos epoch photometry annex (\cite{hipparcos}) and the resulting time series can be found in Figure \ref{fig1}. This data was obtained during the 4 year Hipparcos mission and like the SMEI data shows no significant variability of the source.

\section{Conclusions}

Based on the 3 years of continuous SMEI coverage and the past Hipparcos observations we are able to determine that there appears to be no significant variability in the optical emission from $\alpha$ Cassiopeiae. Past observations have suggested that the star is variable over periods of days. We do not detect any variability on these timescales. Of course we are comparing observations over a century apart and it is intriguing that the variability has since ceased. If the star was genuinely variable and now has ceased then this is an interesting possibility for stellar evolution. Possibly the star experiences variability over larger time scales than our data-set and as such is a perfect target for amateur observers.

\acknowledgments
The authors acknowledge the support of STFC. SMEI was designed and built by members of UCSD, AFRL, and the University of Birmingham.

\end{document}